\documentclass[prl,aps,10pt,showkeys,nofootinbib,showpacs,twocolumn]{revtex4-2}

\usepackage{amsmath,amssymb,amsfonts,amsthm}
\usepackage{physics}
\usepackage{xcolor}
\usepackage{graphicx}
\usepackage{hyperref}

\newcommand{\oarX}[1]{\href{http://arxiv.org/abs/#1}{{\ttfamily #1}}}
\newcommand{\arX}[1]{\href{http://arxiv.org/abs/#1}{{\ttfamily arXiv:#1}}}

\def\be{\begin{equation}}
\def\ee{\end{equation}}
\def\im{{\rm i}}

\begin{document}

\title{Black hole singularity resolution in unimodular gravity from unitarity}

\author{Steffen Gielen}
\affiliation{School of Mathematical and Physical Sciences, University of Sheffield, Hicks Building, Hounsfield Road, Sheffield S3 7RH, United Kingdom}
\email{s.c.gielen@sheffield.ac.uk}
\author{Luc\'ia Men\'endez-Pidal}
\affiliation{Departamento de F\'isica Te\'orica, Universidad Complutense de Madrid, Parque de Ciencias 1, 28040 Madrid, Spain}
\email{lumene02@ucm.es}
\date{\today}


\begin{abstract}
We study the quantum dynamics of an interior planar AdS (anti--de Sitter) black hole, requiring unitarity in the natural time coordinate conjugate to the cosmological ``constant of motion'' appearing in unimodular gravity. Both the classical singularity and the horizon are replaced by a non-singular highly quantum region; semiclassical notions of spacetime evolution are only valid in an intermediate region. For the singularity, our results should be applicable to general black holes: unitarity in unimodular time always implies singularity resolution.
\end{abstract}


\maketitle

{\em Introduction.} --- The fate of classical singularities is one of the most important questions for any theory of quantum gravity; indeed, the incompleteness of classical relativity formalised in the Penrose--Hawking singularity theorems \cite{singtheorem} is often cited as a main motivation for why gravity must be quantum. The most important singularities of direct relevance to our Universe are at the Big Bang and at the centre of black holes. In a first approximation, these situations can be represented through idealised, spatially homogeneous geometries whose high degree of symmetry allows for a quantum description at least at an effective level. One can then ask in such simple models what happens to the classical singularity.

In the context of homogeneous cosmology, the question of whether singularities are resolved through quantum effects (in what is usually called quantum cosmology) does not have a clear answer, since it depends on the criteria for singularity resolution and on the precise definition of the model including the choice of quantum state \cite{QSingRes}. Nevertheless, one can make general statements if the quantum theory is required to be unitary with respect to a given choice of time \cite{GotayDemaret}: one would expect singularities to be resolved if they are only a finite ``time'' away, since singular evolution seems incompatible with requiring a global time translation operator. Inevitably, such a general result means that the property of singularity resolution depends on the choice of clock \cite{SteffenLuciapapers}, and signifies a clash between general covariance and unitarity \cite{Essay}, itself a somewhat controversial topic in quantum gravity. 

Here we note that the results of Ref.~\cite{SteffenLuciapapers} extend straightforwardly to the study of black hole singularities, in particular for the planar AdS black holes studied in Ref.~\cite{HartnollWdW} and related previous work in the context of AdS/CFT \cite{Frenkel}. The interior metric studied in these works is of Kasner type, with a single anisotropy variable, and dynamically equivalent to a flat homogeneous, isotropic cosmology with a massless scalar field. The cosmological constant is taken to be negative and fixed, but reinterpreting it as a constant of motion as suggested by unimodular gravity \cite{unimodular} turns it into another global degree of freedom, conjugate to unimodular time. The black hole interior is then classically identical to the cosmology studied in \cite{SteffenLuciapapers}, and its canonical quantisation can be studied along the same lines. If the Schr\"odinger-like unimodular time is used to define evolution and one requires the theory to be unitary, all singularities are resolved \cite{GrybTheb}. 

Unimodular gravity refers to modifications of general relativity in which only the trace-free Einstein equations are imposed. The cosmological constant becomes an integration constant, and there is one additional global degree of freedom. Classically unimodular gravity and general relativity are indistinguishable; there may be differences in the quantum regime \cite{NewUnimodular}. While we focus on unimodular gravity, our mechanism for singularity resolution could be applied to other extensions of general relativity that include clocks to define unitarity, such as the vacuum sequester \cite{Sequester1} in its local version \cite{Sequester2}, models with additional matter fields \cite{DustModels}, or with constants of Nature becoming constants of motion \cite{Joao}.

The notion of ``singularity'' in this context does not only apply to curvature singularities; a singularity is any point at which the classical evolution terminates, and where  non-classical behaviour is  required for a unitary quantum theory. These can be coordinate singularities \cite{Essay}. In our black hole model, at the horizon the spatial volume goes to zero and the classical solution cannot continue. Quantum unitarity with respect to the preferred clock (unimodular time) would then require the horizon to be similarly replaced by a highly quantum region in which classical evolution is not valid.  This specific conclusion is sensitive to having a foliation that becomes singular at the horizon. However, the conclusion regarding black hole singularities is more generally valid since the singularity is only a finite time away for many observers. The Belinski--Khalatnikov--Lifshitz (BKL) conjecture states that approach to a generic singularity is described by Kasner-like dynamics, like the example studied here; this suggests that for many clocks the classical singularity would need to be replaced by well-defined quantum evolution leading to the emergence of a white hole. Our results demonstrate that black hole singularities either lead to a failure of quantum unitarity (in unimodular time), or to a new scenario for a quantum transition of a black hole to a white hole.

{\em Quantum theory of black hole interior.} --- The classical action for general relativity with cosmological constant $\Lambda$, including the Gibbons--Hawking--York boundary term, is 
\be
S=\frac{1}{\kappa}\int{\rm d}^4 x\sqrt{-g}\left(\frac{1}{2}R-\Lambda\right)-\frac{1}{\kappa}\int {\rm d}^3 x\sqrt{q}K
\label{action}
\ee
where $g_{\mu\nu}$ is the spacetime metric, $R$ the Ricci scalar, $q_{ij}$ the boundary metric and $K$ the extrinsic curvature; $\kappa=8\pi G$ where $G$ is Newton's constant.

The interior planar black hole (Kasner) metric studied in Ref.~\cite{HartnollWdW} and previous papers \cite{Frenkel} is given by
\be
{\rm d}s^2 = -N^2\,{\rm d}r^2 + v^{2/3}\left(e^{4k/3}{\rm d}t^2+e^{-2k/3}({\rm d}x^2+{\rm d}y^2)\right)
\label{metricansatz}
\ee
where $N$, $v$ and $k$ are functions of $r$ only. Thought of as a radial coordinate outside the horizon, in the interior $r$ is timelike and hence this metric is spatially homogeneous. It corresponds to a locally rotationally symmetric Bianchi I model with metric written in the Misner parametrisation (see, e.g., Ref.~\cite{bojobook}). One important feature of this parametrisation is that the anisotropy variables (here a single one, $k$) behave as free massless scalar fields in a flat isotropic geometry, as we will see explicitly.

Substituting the metric ansatz (\ref{metricansatz}) into (\ref{action}) reduces the action to
\be
S = \frac{1}{\kappa}\int_{-\infty}^\infty {\rm d}r\, \left(\frac{(k')^2 v^2 - (v')^2}{3Nv}-\Lambda N v\right)
\label{MSSaction}
\ee
where $'$ denotes derivative with respect to $r$. We have implicitly assumed that the overall numerical factor arising from performing the integration over $t$, $x$ and $y$ has been set to one by a choice of coordinates.

Legendre transform leads to a Hamiltonian
\be
\mathcal{H}=\frac{\sqrt{3}}{2}N\,v\left(\frac{\pi_k^2}{v^2}-\pi_v^2+\Lambda\right)\,.
\ee
where we have made the unit choice $\kappa=2/\sqrt{3}$ to obtain a simpler form.
The resulting Hamiltonian constraint \cite{HartnollWdW}
\be
\mathcal{C}=\frac{\pi_k^2}{v^2}-\pi_v^2+\Lambda\approx 0
\label{constraint}
\ee
is exactly the one studied in Ref.~\cite{SteffenLuciapapers}, where it was interpreted as describing a flat homogeneous, isotropic cosmology with a free massless scalar field and a perfect fluid (playing the rôle of dark energy). In this interpretation, $\Lambda$ is no longer a parameter but a (conserved) momentum conjugate to a time variable $T$. This assumption can be justified by promoting $\Lambda$ in (\ref{MSSaction}) to a dynamical variable and adding a term $\Lambda\,T'$ to the action. The resulting action is then the reduction of the Henneaux--Teitelboim action for unimodular gravity \cite{HenneauxTeitelboim}
\be
    S_{{\rm HT}}=\frac{1}{\kappa}\int {\rm d}^4 x\left[\sqrt{-g}\,\left(\frac{1}{2}R-\Lambda\right)+\Lambda\partial_\mu\mathcal{T}^\mu\right]
\ee
with suitable boundary terms to a spatially homogeneous geometry. Similar constructions would be possible for other theories as mentioned above.

The classical solutions in the time $T$ are found to be 
\begin{align}
v(T)& = \sqrt{-\frac{\pi_k^2}{\Lambda}+4\Lambda(T-T_0)^2}\, ,
\label{vsolution}
\\k(T) &= \frac{1}{2}\log\left|\frac{\pi_k-2\Lambda(T-T_0)}{\pi_k+2\Lambda(T-T_0)}\right| + k_0
\end{align}
where $T_0$ and $k_0$ are integration constants. The metric \ref{metricansatz} has singularities ($v\rightarrow 0$) for $T_-=T_0-\frac{\pi_k}{2\Lambda}$ and $T_+=T_0+\frac{\pi_k}{2\Lambda}$. The Kretschmann scalar
\be
R_{\mu\nu\xi\eta}R^{\mu\nu\xi\eta}=\frac{8\Lambda^2}{3}\left(1+\frac{2\pi_k^2}{\left(2\Lambda(T-T_0)+\pi_k \right)^2} \right)
\label{kretsch}
\ee
diverges for $T\rightarrow T_-$ (black hole singularity, $k\rightarrow +\infty$) but is finite for $T=T_+$ (black hole horizon, $k\rightarrow -\infty$). 

The constraint (\ref{constraint}) can be written as $\mathcal{C}=g^{AB}\pi_A \pi_B+\Lambda$, making the dynamical problem equivalent to relativistic particle motion in a configuration space (minisuperspace) parametrised by $v$ and $k$ and with a flat metric 
\be 
g_{AB}=\begin{pmatrix} v^2 & 0 \cr 0 & -1 \end{pmatrix}\,.
\ee 
This minisuperspace is equivalent to the Rindler wedge, a portion of full (1+1) dimensional Minkowski spacetime with boundary at $v=0$. This viewpoint suggests a natural operator ordering in canonical quantisation \cite{SteffenLuciapapers,HartnollWdW,HawkingPage,MalcolmBH}, leading to the Wheeler--DeWitt equation
\be
\left(\Box+\Lambda\right)\Psi:=\left(-\frac{1}{v^2}\partial_k^2+\partial_v^2+\frac{1}{v}\partial_v+\Lambda \right)\Psi=0
\label{WdWeq}
\ee
where $\Box$ is the Laplace--Beltrami operator for $g_{AB}$. Eq.~(\ref{WdWeq}) is covariant with respect to variable transformations of $v$ and $k$ and, because $g_{AB}$ is flat, with respect to lapse redefinitions which act as conformal transformations on $g_{AB}$ \cite{Halliwell}. 

The general solution to Eq.~(\ref{WdWeq}) can be straightforwardly given as
\begin{align}
\Psi(v,k,\Lambda)&=\int_{-\infty}^\infty \frac{{\rm d}p}{2\pi}\,e^{\im pk}\left(\alpha(p,\Lambda)J_{\im p}\left(\sqrt{\Lambda}v\right)\right.\nonumber
\\&\qquad\left.+\beta(p,\Lambda)J_{-\im p}\left(\sqrt{\Lambda}v\right)\right)
\label{generalsol}
\end{align}
where $\alpha(p,\Lambda)$ and $\beta(p,\Lambda)$ are so far arbitrary and $J_\nu(x)$ is a Bessel function of the first kind. Eq.~(\ref{generalsol}) gives the general solution as a function of $\Lambda$, a dynamical variable in our setup; for $\Lambda<0$ it is less ambiguous to pass from $J_{\im p}\left(\sqrt{\Lambda}v\right)$ to the modified Bessel functions $K_{\im p}\left(\sqrt{-\Lambda}v\right)$ and $I_{\im p}\left(\sqrt{-\Lambda}v\right)$ \cite{SteffenLuciapapers}. The appearance of Bessel function solutions for the black hole interior is familiar from other contexts \cite{AnnihtoNothing}. Fourier transform converts the wavefunction given as a function of $\Lambda$ into a time-dependent wavefunction dependent on $T$, the conjugate to $\Lambda$.

Asking whether the resulting quantum theory is unitary with respect to evolution in unimodular time $T$\footnote{To see why $T$ is unimodular time, note the global factor $N\, v=\sqrt{-g}$ in the Hamiltonian $\mathcal{H}$. For $\mathcal{H}$ to be given just by the constraint $\mathcal{C}$, we must choose $N$ so that $\sqrt{-g}$ is constant.} is equivalent to asking whether $\Box$ is self-adjoint in a suitable inner product. The most natural choice of inner product is induced by $g_{AB}$,
\be
\langle \Psi|\Phi \rangle = \int_0^\infty {\rm d}v\;v \int_{-\infty}^\infty {\rm d}k\;\Psi^*\,\Phi\,.
\ee
It is easy to see $\Box$ is then {\em not} self-adjoint, as expected for a space with boundaries which can be reached by classical solutions in a finite amount of time. Here this corresponds to both the black hole horizon and the singularity being only a finite (unimodular) time away. Self-adjoint extensions can be defined by restricting wavefunctions to a subspace satisfying the boundary condition \cite{SteffenLuciapapers,GrybTheb}
\be
\lim_{v\rightarrow 0}\int_{-\infty}^\infty {\rm d}k\;v\left(\Psi^*\frac{\partial\Phi}{\partial v}-\Phi\frac{\partial\Psi^*}{\partial v}\right)=0\,,
\label{bcond}
\ee
thus obtaining a unitary quantum theory. Such a boundary condition can be seen as reflection from the singularity, similar to what is proposed in Ref.~\cite{MalcolmBH}. 

We are interested in $\Lambda<0$ solutions, which are the analogue of bound states. Normalised solutions to the Wheeler--DeWitt equation \ref{WdWeq} and the boundary condition \ref{bcond} can be expressed as
\begin{align}
\Psi(v, k, T)&=\int_{-\infty}^\infty \frac{\dd p}{2\pi} e^{\im p k} \sum_{n\in\mathbb{Z}} e^{\im \Lambda_n^p T}\sqrt{\frac{-2\Lambda_n^p\sinh(\abs{p}\pi)}{\abs{p}\pi}} \nonumber \\
&\alpha(p,\Lambda_n^p)K_{\im \abs{p}}\left( \sqrt{-\Lambda_n^p} v\right)\, ,
\label{normsol}
\end{align}
where $\int \frac{\dd p}{2\pi} \sum_{n\in\mathbb{Z}}\abs{\alpha(p,\Lambda_n^p)}^2=1$ and
\begin{equation}
\Lambda_n^p =- e^{-\frac{(2n+1)\pi}{\abs{p}}+\theta(p)}
\end{equation}
is a discrete set of allowed negative $\Lambda$ values. Here the free function $\theta(p)$ parametrises different self-adjoint extensions of $\Box$; in a sense, each choice of $\theta(p)$ defines a different theory. Physically, $\theta(p)$ encodes a phase shift as the modes are reflected off $v=0$; an extensive discussion can be found in Ref.~\cite{GrybTheb}. We are interested in qualitative features of a theory which replaces the black hole singularity and horizon by non-singular quantum regions, which do not depend on the detailed choice. This insensitivity was also studied in detail in the same model (with a different clock) in Ref.~\cite{Pawlowski2012}; different self-adjoint extensions show only small quantitative differences. In the following we choose $\theta(p)=\pi/|p|$.

Eq.~\ref{normsol} represents the quantum states of a planar AdS black hole interior as superpositions of different values of momentum $p$ and cosmological constant $\Lambda_n^p$. Crucially, because of the reflecting boundary condition the allowed bound states are modified Bessel functions of the second kind, behaving near $v=0$ as
\be
K_{\im \abs{p}}\left( \sqrt{-\Lambda_n^p} v\right) \sim \frac{\Gamma(-\im p)}{2}e^{\im p \log(\frac{\sqrt{-\Lambda_n^p} v}{2})} + {\rm c.c.}\,,
\ee
i.e., as superpositions of positive and negative $p$ modes with equal magnitude. Since changing the sign of $p$ is equivalent to time reversal, swapping the roles of horizon and singularity, or switching between classical black-hole and white-hole solutions (see Eq.~(\ref{kretsch}) and below; this corresponds to the parameter $\pi_k$ in the classical solution), none of these bound states can correspond to a single semiclassical trajectory that ends in a singularity. The necessary superpositions of black-hole and white-hole solutions then lead to singularity resolution in this theory.

\emph{Singularity resolution.} --- To see this explicitly, we numerically study the evolution of a semiclassical state
\begin{equation}
\alpha(p,\Lambda_n^p)=\sqrt{\mathcal{N}} e^{-\frac{(p-p_0)^2}{2\sigma_p^2}-\frac{(\Lambda_n^p-\Lambda_0)^2}{2\sigma_\Lambda^2}}
\end{equation}
with free parameters $p_0$, $\Lambda_0$, $\sigma_p^2$, and $\sigma_\Lambda^2$ and a normalisation factor $\mathcal{N}$. For a semiclassical interpretation we need $\sigma_p\ll p_0$, $\sigma_\Lambda\ll|\Lambda_0|$ and $p_0\gg 1$. The latter condition then also guarantees that the allowed discrete $\Lambda_n^p$ values are reasonably close together. 

Our main observable is the volume $v(T)$. Expectation values and moments of $v$ in our state take the form
\begin{align}
\langle v^\alpha(T)\rangle &= \mathcal{N}\int_0^\infty {\rm d}v\;v^{\alpha+1} \int_{-\infty}^\infty \frac{{\rm d}p}{2\pi}\sum_{n,n'} \frac{2\sqrt{\Lambda_n^{p}\Lambda_{n'}^{p}}\sinh(|p|\pi)}{|p|\pi} \nonumber
\\& \quad\times e^{-\frac{(p-p_0)^2}{\sigma_p^2}}e^{-\frac{(\Lambda_n^{p}-\Lambda_0)^2}{2\sigma_\Lambda^2}-\frac{(\Lambda_{n'}^{p}-\Lambda_0)^2}{2\sigma_\Lambda^2}+\im(\Lambda_n^{p}-\Lambda_{n'}^{p})T}\nonumber
\\& \quad\times K_{\im |p|}\left(\sqrt{-\Lambda_n^{p}}v\right)K_{\im |p|}\left(\sqrt{-\Lambda_{n'}^{p}}v\right)\,;
\label{integral}
\end{align}
the $v$ integral in Eq.~(\ref{integral}) can be done analytically, leaving the sums and the $p$ integral for numerical evaluation. Due to the Gaussians inside the integral, the contributions from $|p-p_0|\gg \sigma_p$ and $|\Lambda_n^p-\Lambda_0|\gg\sigma_\Lambda$ are very small and we can replace the infinite $p$ integral and sums over $n$ and $n'$ by finite sums by introducing cutoffs. We chose these so that the integrals and sums include the regions $\abs{p-p_0}\leq 5 \sigma_p$ and $\abs{\Lambda_p^n-\Lambda_0}\leq 5 \sigma_\Lambda$; increasing the cutoff does not lead to any noticeable improvement.

The expectation value $\langle v(T)\rangle$ can be compared with classical solutions given in Eq.~(\ref{vsolution}) where $\pi_k$ and $\Lambda$ are replaced by the average of $p$ and $\Lambda$ in our chosen states; due to the discrete spacing of possible $\Lambda$ values these averages are not exactly equal (but close) to $p_0$ and $\Lambda_0$.

\begin{center}
\begin{figure}[htp]
\includegraphics[scale=0.9]{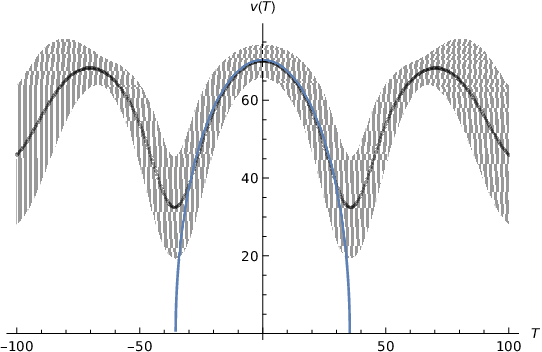}
\includegraphics[scale=0.9]{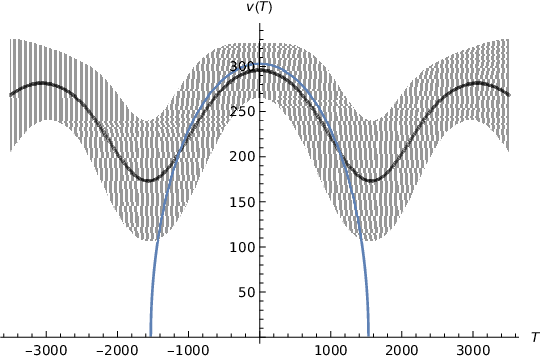}
\caption{Quantum expectation values $\langle v(T)\rangle$ (400 black data points) with fluctuations (in grey) for a state with $p_0=70$, $\sigma_p=0.1$, $\Lambda_0=-1$ and $\sigma_\Lambda=0.15$ (top) and one with $p_0=30$, $\sigma_p=0.05$, $\Lambda_0=-0.01$ and $\sigma_\Lambda=0.0019$ (bottom), compared with the corresponding classical solution (\ref{vsolution}) (in blue). The points where classically $v\rightarrow 0$ correspond to the horizon and the singularity. }
\label{BHplots}
\end{figure}
\end{center}

In Fig.~\ref{BHplots} we show the quantum expectation value $\langle v(T)\rangle$ and fluctuations $\Delta v=\sqrt{\langle v(T)^2\rangle - \langle v(T)\rangle^2}$ for such a state \cite{Data}. We can see that, as expected, for small $|T|$ the expectation value stays close to its corresponding classical solution (\ref{vsolution}), but it departs strongly near the horizon or singularity where interference between an ingoing (black hole) and outgoing (white hole) solution becomes important. There is a finite minimal value for $v$ and in this sense, both singularity and horizon are replaced by quantum ``bounces''. Where the expectation value closely follows the classical curve the variance is small, but at the bounces the variance grows, indicating strong quantum fluctuations where the state is reflected. As required by unitarity, all expectation values and higher moments are globally defined, not just for the finite $T$ interval in which the classical solution is valid. Taken at face value the resulting quantum solution describes cycles of local expansion and contraction, corresponding to a sequence of black hole/white hole interiors passing from horizon to singularity and back. Over longer timescales the variances grow, suggesting a spreading in the state and eventual breakdown of the semiclassical picture. While all the specific features displayed here depend on the chosen parameters in the state, the qualitative behaviour showing disappearance of the classical horizon and singularity seems universal, resulting from the reflecting boundary condition (\ref{bcond}). 

{\em Discussion.} --- It has long been stated that a quantum theory of black hole dynamics that is required to be unitary must deviate strongly from semiclassical expectations. Usually this is discussed in the context of unitarity of black hole formation and evaporation, leading to the famous issue of information loss \cite{HawkingBH}. Preserving unitarity together with a few other ``reasonable'' assumptions can lead to disasters such as a firewall at the horizon \cite{Firewall}, or more generally the conclusion that there is no simple semiclassical resolution of the paradox.  What we are discussing here is different; we studied a simple quantum model of the black hole interior, in which the gravitational degrees of freedom (truncated to a Kasner metric) are quantised using the Wheeler--DeWitt equation. In this setting too, there is a clash between unitarity and consistency with semiclassical physics: unitarity means globally well-defined time evolution which is incompatible with singularities, even a coordinate singularity at the horizon. It is a tricky issue to define unitarity in a fundamentally timeless setting such as canonical quantum gravity. The key ingredient in our discussion was to use unimodular gravity and its preferred choice of time, implemented through auxiliary fields as in Ref.~\cite{HenneauxTeitelboim}. With respect to this clock, both the horizon and the singularity are only a finite time away, as they would be for an infalling observer. Unitarity forces us to replace the horizon and singularity with highly quantum bounce regions. Unitarity with respect to a different clock would generally lead to different conclusions \cite{Essay}. Our general conclusions should be valid for {\em any} standard of time such that the singularity or horizon are only a finite time away; it would be interesting to construct other analytically accessible examples. Again we point out that this could be done in other theories that extend general relativity by including local clock variables \cite{Sequester1,Sequester2,DustModels,Joao}.

The work of Ref.~\cite{HartnollWdW} studied a number of clocks in the same interior black hole spacetime (see also Ref.~\cite{BlackerSirui} for an extension to charged black holes). For instance, one can use the classically monotonic anisotropy parameter $k$. With respect to $k$ or other clocks such as $v$ and $\pi_v$, no deviation from semiclassical physics was found in Ref.~\cite{HartnollWdW}. These observations are fully consistent with the results of Ref.~\cite{SteffenLuciapapers} for $k$ and $v$, and with the general conjecture of Ref.~\cite{GotayDemaret}, where different clocks were classed as ``slow'' and ``fast''. Unitarity with respect to a fast clock, which runs to $\pm\infty$ at a singularity,  does not require any deviation from semiclassical physics. In our model $k$, $v$ and $\pi_v$ are all fast. Similar behaviour is also found for a massless scalar field in homogeneous Wheeler--DeWitt cosmology \cite{AshtekarSingh}. However, such clocks do not describe the experience of local observers; classical singularities are troublesome exactly because one can reach them in finite time. When such a slow clock is studied, on approach to the singularity one must either give up unitarity or find generic resolution of singularities and possibly horizons.

The metric form (\ref{metricansatz}) corresponds to a simple model of a black hole with planar symmetry, but only a slight extension -- adding a second possible anisotropy variable in the Misner parametrisation -- turns it into the general Kasner form describing, according to the BKL conjecture, successive periods during the generic approach to a spacelike singularity, even for more realistic black holes (see also the related Ref.~\cite{MalcolmBH}). Since this second anisotropy variable again acts as a massless scalar field in an isotropic Universe, the results illustrated here would be expected to hold more generically for singularities. For horizons, the general picture is less clear since the model studied here sees the horizon as a coordinate singularity, and the black hole metric at a horizon is in general more complicated. Already for the usual (A)dS-Schwarzschild black hole, the positive curvature of constant time slices would contribute at the horizon and potentially change the conclusions. In general, it is in principle always possible to construct a notion of time that stays regular at the horizon, so that collapse from an asymptotically flat (or asymptotically AdS or de Sitter) region could be described as a unitary process. While these alternative constructions will change the interpretation of the horizon, in the theory we have defined unitary quantum dynamics will always necessarily replace the universal singularity by non-singular evolution into a white hole: either unitarity fails, or there is no black hole singularity.

\emph{Acknowledgments.} --- The work of SG is funded by the Royal Society through the University Research Fellowship Renewal URF$\backslash$R$\backslash$221005. LMP is supported by the Leverhulme Trust.


\begin{thebibliography}{99}
\bibitem{Data}
The supporting data for this article (in the form of a {\em Mathematica} notebook) are openly available from the {\tt arXiv} version of the paper, \arX{2409.03006}.
\end{thebibliography}
\end{document}